# Slow waves in locally resonant metamaterials line defect waveguides

Nadège Kaina[1], Alexandre Causier[2], Yoan Bourlier[2], Mathias Fink[1], Thomas Berthelot[2], Geoffroy Lerosey[1]

[1] Institut Langevin, ESPCI ParisTech & CNRS, Paris, France
[2] CEA Saclay, IRAMIS, NIMBE, LICSEN, UMR 3685, F-91191, Gif sur Yvette, France

*In the past decades, many efforts have been devoted to the temporal manipulation of waves, especially focusing on slowing down their propagation. In electromagnetism, from microwave to optics, as well as in acoustics or for elastic waves, slow wave propagation indeed largely benefits both applied and fundamental physics. It is for instance essential in analog signal computing through the design of components such as delay lines and buffers, and it is one of the prerequisite for increased wave/matter interactions. Despite the interest of a broad community, researches have mostly been conducted in optics along with the development of wavelength scaled structured composite media, that appear promising candidates for compact slow light components. Yet their minimum structural scale prevents them from being transposed to lower frequencies, such as terahertz or microwave in electromagnetism, ultrasound or audible in acoustics, where wavelengths range from sub-millimeter to meters. In this article, we propose to overcome this limitation thanks to the deep sub-wavelength scale of locally resonant metamaterials. In our approach, implemented here in the microwave regime, we show that introducing coupled resonant defects in such composite media allows the creation of deep sub-wavelength waveguides. We experimentally demonstrate that waves, while propagating in such waveguides, exhibit largely reduced group velocities. We qualitatively explain the mechanism underlying this slow wave propagation and first experimentally demonstrate, then numerically verify, how it can be taken advantage of to further tune the velocity, achieving group indices $n_g$ as high as 227 over relatively large bandwidths. We conclude by highlighting the three extensively beneficial consequences of our line defect slow wave waveguides in locally resonant metamaterials. On the one hand, the deep sub-wavelength scale of the waveguides makes it a practical compact platform for low frequencies. Then, it provides very large group indices that together with the extreme field confinement are the two elementary requirements to efficiently enhance the probability of wave/matter interactions. Finally and interestingly, on the contrary to other approaches of the state of the art, slow wave propagation in such metamaterial waveguides does not occur at the expense of drastic bandwidth reductions.*

Being able to temporally control the propagation of waves, and particularly achieving low group velocities, is one of the current challenges in wave physics, regarding its many related outcomes in both applied and fundamental physics. First, there are obvious benefits for the development of devices and the simplification of information processing through the design of delay lines and buffers. From a more fundamental point of view, slow waves are of primary importance to observe and exploit many fascinating, though weak, effects, of either quantum or classical physics. It indeed permits to enhance the wave-matter interaction probability by extending the interaction time in the propagation medium. Most commonly studied in the optical field, those interactions can for instance benefit low-power nonlinear physics[1], efficient solar energy harvesting[2], miniature lasers[3] or be used to control the spontaneous emission rate of atoms[4]. However, lower frequency wave-matter interactions are also relevant and hence widely studied, going from quantum QED[5], masers[6,7], electronic spin transition in NV centers[8,9] or nonreciprocal devices. Moreover, the interest of slow wave propagation in applied physics is not the prerogative of short wavelengths since it finds numerous applications at lower frequencies, from the near infrared to microwaves, as in the fields of elastic waves and phononics. It is indeed required to enhance the efficiency of sensors, for the detection of hazardous products, for the development of phononic based circuits[10] or for analog processing of microwaves in telecommunications[11], which actually often relies on modulations to optical frequencies[12] hence resulting in very high insertion losses.

It is nonetheless in optics, thanks to the extensive researches on structured based composite materials, namely photonic crystals[13], that most of the components were developed. Those Bragg interference based media are indeed considered as natural candidates due to their ability to support line defect waveguides, named photonic crystal waveguides, or defect cavities which can be weakly coupled to create photonic crystal cavity waveguides[14,15], both leading to slow wave propagation[16–18] though implying different mechanisms. While the first ones rely on the dispersion induced in the waveguide[19], the second, that are one particular implementation of coupled resonators optical waveguides (CROWs), exploit the strong dispersion brought by the resonances of its elementary constituents, namely, cavities[20–22].

Yet the use of these Bragg interference based media or CROWs sets several constraints, leading to two fundamental limitations. First, the overall dimensions of related components, as well as the wave spatial confinement, are set by the wavelength. Indeed, photonic crystal properties mainly rely on their structure, which is by definition wavelength scaled, as are the dimensions of elementary resonators in CROWs. Hence, if they remain reasonably small in optics, they obviously cannot be used as compact components at lower frequencies, as in the microwave domain for instance where they would end up on meter scaled components. Second, in both systems, slow wave propagation always comes at the expense of a bandwidth narrowing[23], coming either as a less trivial consequence of the wavelength scale of photonic crystal or the use of extremely resonant unit cells (equivalently very high quality factor resonators) for CROWs.

In this article, we investigate how both of these constraints can be released by transposing the previous slow wave concepts into the metamaterial field. Metamaterials, composite media generally composed of resonators organized at deep-subwavelength scales, indeed present effective properties that can be very high, a property that was used to image or focus waves below the diffraction limit[24–30], near-zero[31] or negative[32,33]. In a previous work[34], we evidenced that metamaterials negative effective properties can also be understood as bandgaps that originate from the interference between the plane waves propagating in the host matrix and the fields radiated by their resonant elements. This purely interference based interpretation of the origin of the metamaterials bandgap, that emphasizes analogies with the physics of photonic crystals, allowed us to demonstrate that local modifications of the material, analogous to photonic crystals, induce defect cavities able to confine waves to the defect region. Contrarily to structured composite media however, for which the bandgap is created by interferences on the periodic structure, and hence the defects are structural ones with sizes the order of the wavelength, the hybridization bandgap in metamaterial mainly stems from the resonant nature of its constituents[35,36]. Defect cavities then typically consist in modifying the resonance frequency of one element so that it falls within the hybridization bandgap of the surrounding medium. It results in wave confinements on much smaller dimensions and in very high Purcell factors[37], due to the typical deep sub-wavelength metamaterial scales. Following this idea, we further proved that inserting sequences of those resonant defects enables the creation of waveguides that can mold the flow of waves, again at deep subwavelength scales[34,38]. Through this initial proof of concept study, it was shown that the first constraint regarding the scale of components can be largely relaxed using metamaterials for which defect cavities and waveguides dimensions are independent of the wavelength.

Here, we concentrate on the temporal properties of such line defect waveguides in locally resonant metamaterials. While experimentally measuring the delays of pulses propagating in the waveguides, we first prove that the latter can slow down wave very effectively, with experimental group indices as high as $n_g = 227$, challenging the performances of photonic crystal based delay lines. We show that this value is furthermore largely tunable through modifications of the geometric properties of the surrounding bandgap medium. Moreover, recording both the transmission spectra and the delays enables us accessing the essential figure of merit that characterizes the efficiency of any slow wave device: the normalized delay-bandwidth product (NDBP), a dimensionless number independent of the length of the device that underlines how broadband the latter can operate given its slowing 'strength', i.e. its group index $n_g$. It is defined as NDBP= $v_g/c*\Delta f/f_0 = n_g*\Delta f/f_0$, where $\Delta f$ is the bandwidth and $f_0$ the central frequency of the device. In other composite media based slow wave devices, this NDBP is intrinsically limited: the slower the wave, the narrower the bandwidth[39]. Though efforts have been made[19,40] to overcome this limitation, NDBP typical values remain less than unity. Strikingly, we observe in the defect line waveguides in metamaterials NDBPs at least one order of magnitude higher than proposals from the state of the art, coming mostly from the photonic crystal community[41]. More importantly, we show that, owing to their deep subwavelength scale, the NDBP in locally resonant metamaterials is no longer limited but can be easily increased by tailoring the density of the metamaterial line defect waveguides. This in turn overcomes the second usual intrinsic limitation of slow wave devices. We confirm this property via simulations and prove that the NDBP evolves as the inverse of the metamaterial typical spatial scale, independently of the wavelength, a property unique to the resonance driven physical mechanism of these media. Finally, we conclude by highlighting the three essential consequences of using line defect waveguides in metamaterials to slow down microwaves, namely deep sub-wavelength confinement, large group indices and extended bandwidth, while discussing how these can be of interest for both device implementations and wave-matter interactions, especially for low frequency waves.

**RESULTS**

**Measuring the spectral, spatial and temporal properties of a metamaterial line defect waveguide**. To support our demonstration, we experimentally and numerically study microwave samples based on arrays of quarter-wavelength resonant metallic wires, of given height L, sitting on a ground plane. For a defect created by shortening a wire to a length $L_d$ (or equivalently increasing its resonance frequency[34]), this collection of resonators acts as a bandgap medium. By shortening several adjacent wires along a line, a metamaterial line defect waveguide is inserted, for which we measure both the temporal and spectral properties. To first explore the overall characteristics of the propagation in such metamaterial line defect waveguides, we manufacture a 3D printed periodic subwavelength square array (period a = 5 mm ~ $\lambda_0/13$) of L = 16 mm long copper-plated wires mounted on a ground plane (see SI for the 3D printing and copper-plating process). Such wires have a resonance frequency close to $f_0$ = 4.5 GHz, so that this medium opens a bandgap above $f_0$[34]. The waveguide is then designed by replacing some of the wires along a tortuous path with $L_d$ = 14 mm long defect wires, as depicted in Figure 1a. To quantitatively measure the waveguide properties, two homemade input and output antennas, each one plugged to a network analyzer, feed the latter. Those antennas are designed to be impedance matched to the defect wires to maximize the coupling to the waveguide and avoid reflections (see SI). The acquired spectrum (Fig. 1c) first highlights the sudden drop of transmission above $f_0$, corresponding to the bandgap created by the long wires. It then evidences a transmission band of 6% bandwidth (Δf = 300 MHz) around a central frequency $f_c$ = 4.88 GHz, close to the defects resonance. This band can be attributed to the presence of the line defect waveguide, as the spectral transmission of an array of all identical wires only displays a bandgap for this frequency range[34] (see SI). Note that for this band, the transmission spectrum is not completely flat, as we could have expected using impedance matched antennas. This is due to some small reflections that are experienced by the wave around the numerous corners of this meandering propagation path, that can however be minimized by adapting the waveguide geometry around the bends, as in photonic crystal waveguides[42]. To further assess that the defect line indeed acts as a deep-subwavelength waveguide, we record the electric field by scanning the medium over a wide frequency range with a homemade antenna mounted on a 2D moving stage, while the antenna at the output port of the waveguide is plugged in to a standard 50 Ω load (see SI). A map of one of the modes within the defects transmission band (Fig. 1b) highlights the effective deep subwavelength spatial confinement of the electric field around the defect wires (white circles). We see that the confinement is limited by the distance to the first neighbors in the surrounding medium (black circles) corresponding to twice the lattice constant, i.e. $\lambda_0/6.5$. This result confirms our first experimental proof of concept[34] while further demonstrating two things: on the one hand waves can be molded along arbitrary complex paths whereas on the other hand the coupling of adequate feeding antennas to the defect line waveguide ensures one of the essential prerequisite to turn it into an effective component. Finally, the temporal properties (the delay or equivalently the group velocity) of the microwaves propagating in the waveguide can be estimated. To do so, a short pulse is first sent in the input antenna and received at the output antenna (Fig. 1d, up), showing a strong delay of about 20 times the FWHM of the sent pulse though the waveguide is only a few wavelengths long. To quantitatively measure the group velocity, which is dispersive, we then choose to filter this pulse on a 20 MHz wide bandwidth centered on $f_c$ (Fig. 1d, down). While logically artificially broadening the signals, the filtering operation allows to evidence that the output pulse is received after a 55 ns delay that can be attributed to the propagation through the $L_g$ = 29.5 cm long meandering path. This delay, equivalent to a propagation on 16.5 m in air, corresponds to a group index $n_g$ = 56, and hence demonstrates that waves are strongly slowed down in the metamaterial defect waveguide, which is what we will focus on from now on.

**Origin of the slow wave propagation in metamaterial line defect waveguides**. To understand the origin of this very high group index and determine if it can be further enhanced, it is necessary to understand the physics underlying the propagation of waves in these line defect waveguides. The shorter wires forming the waveguide are resonant defects within the bandgap created by the resonance of the longer wires constituting the surrounding medium. The field is then necessarily tightly confined around each defect. The subwavelength transverse size of the waveguide furthermore proscribes the presence of propagating waves. Each of these defects can hence only be, in first approximation, evanescently coupled to its neighbors. Waves then tunnel from defect to defect through a tight binding like interaction, characterized by a coupling strength $\kappa$. This mechanism is very similar to that of CROWs in optics[43], which have been experimentally realized using ultra-high Q cavities such as ring resonators[44,45] or defect cavities in photonic crystals[15,46]. Yet metamaterials are structured at deep subwavelength scales, much smaller than that of CROWs whose resonant elements can be several wavelengths large, or even that of photonic crystals waveguides, which are typically wavelength scaled. We will see later on that this feature considerably enhances the slowing properties of these waveguides and especially their corresponding NDBP. Knowing the propagation mechanism in the line defect waveguide, we now have a clear insight on how to further increase the group index $n_g$, because the group velocity is partly driven by the coupling strength $\kappa$. Indeed, the lower the coupling, i.e. the more difficult the tunneling, the slower the propagation and the higher the group index $n_g$. The idea is then simply to modulate $\kappa$, which can be done while acting on one of the two characteristics affecting the tunneling efficiency: the distance between defects (i.e. the period in the waveguide) or the field confinement around each defect. The latter is in practice set by the bandgap attenuation efficiency that can easily be modified by adjusting the density of resonators in the surrounding medium. The denser the bandgap medium is, the more efficient the bandgap attenuation becomes (Fig. 2a). The field around the defects in the waveguide is then more tightly confined, decreasing the coupling strength $\kappa$ and resulting in smaller group velocities. Note that such a modification of the surrounding medium's structure is a unique prerogative offered by metamaterials, since they are subwavelength scaled and governed by a local resonance rather than by Bragg interference as in structured materials. The latter indeed have a determined periodicity which is constrained by the wavelength and that cannot be freely scaled without scaling the operating frequency as well.

**Tuning $n_g$ through the modification of the defect-defect tunneling strength**. For the sake of simplicity, the group velocity tailoring is now demonstrated in straight waveguides. They are composed of 60 of the previous defect wires separated by a period $a_g$ = 1.5 mm ~ $\lambda_0/40$. The surrounding medium's wires are packed in a periodic array of lattice constant a, varying from 5 mm to 2 mm in the four investigated samples (Fig. 2a), corresponding to periodicities ranging from $\lambda_0/13$ to $\lambda_0/30$. We first implement temporal measurements in order to evaluate the group velocity in our waveguides. For each sample, we send a 20 MHz wide pulse centered on the transmission band central frequency and measure the temporal delay of the received signal (see Fig. S7 in SI for short pulse measurements). The results, presented in figure 2b, evidence that shrinking the medium periodicity from a = 5 mm to a = 2 mm indeed strongly modulates the delay, that extends from 20 ns to 68 ns corresponding to group indices from $n_g$ = 65 to $n_g$ = 227. This strong delay is equivalent to free space propagation of 5.8 m to 20.4 m, although the waveguide physical length is only $L_g$ = 9 cm. Such high $n_g$ is already comparable to the best performances obtained in optics with structured composite materials based slow light devices. Interestingly, the group index presents a linear dependence on the inverse of the medium density $a^2$ (Fig. 2c) and is hence highly tunable and improvable, since the lattice constant is theoretically solely limited by the minimum spacing between the wires, namely their radius. After successfully proving that the group velocity can be tailored, it is important to determine how the transmission bandwidth is affected by the velocity reduction, which is a natural concern regarding slow wave propagation. The measured transmission spectra for each waveguide (Fig. 2d) show that while increasing the group index, the transmission bandwidth accordingly shrinks from $\Delta f$ = 610 MHz to $\Delta f$ = 250 MHz, that is from 10% to 2% of the central frequency. This is not surprising since in a tight-binding model, the bandwidth $\Delta f$ is directly proportional to the coupling strength $\kappa$. As we decreased $\kappa$ to slow down microwaves, through the surrounding medium density, the bandwidth is naturally highly impacted. Note that for this linear geometry of waveguides, the transmission bandwidths are flat, confirming the impedance matching of the antennas.

**Delay-bandwidth products in locally resonant metamaterial-based line defects waveguides**. Despite this bandwidth reduction, one substantial observation is that $\Delta f$ remains very large compared to what can be obtained in conventional composite structure slow wave devices. This is made particularly clear while calculating the corresponding NDBP for each waveguide, ranging from 7.5 (a = 5 mm) to 10.6 (a = 2 mm), values that are more than 10 times higher than any prior proposition based on CROWs or PC. This feature can be explained as a direct consequence of the deep subwavelength nature of metamaterials, as evidenced in Figure 3. We represent here the schematic sinusoidal dispersion relation typical of tight binding interactions that take place in both defect cavities photonic crystal waveguides/CROWs and our line defect waveguides in metamaterials. These dispersion relations inform simultaneously about the group indices (through the slope of the curves, for instance around the central frequency) and about the total bandwidth that can be achieved in both kind of defect chains. To compare the two kind of materials, that are structured media and metamaterials, we consider two basic cases. In the first one (Fig. 3a), both linear chains of defects are supposed to exhibit the same bandwidth $\Delta f$, corresponding to an equivalent coupling strength $\kappa$. As it is pictured, the group index in the case of metamaterial line defect waveguides is then largely enhance compared to the structured material case. This can be attributed to the difference of the first Brillouin zone extension, i.e. the maximum values of wavenumbers that can be reached for waves propagating in the waveguide (for the lower frequency branch we are interested in), in each kind of material. These wavenumbers are indeed only governed by the periodicity in the waveguide, through $\pi/a_g$. In photonic crystal or CROWs structures (blue curve), $a_g$ between the defects is typically of the order of the wavelength while in metamaterials (red curve) this periodicity is deeply subwavelength. Our line defect waveguides can then support much higher wavenumber modes than other structured media based slow wave devices. This implies that for a given bandwidth, the group index can be much higher in metamaterials, which is geometrically observable through the difference of slopes in the dispersion relations close to the central frequency $f_c$ (Fig. 3a). In the second case, we assume that the group indices of both type of chains are now identical (the slopes are the same around the central frequency). Again, the schematic view of Figure 3b demonstrates that the difference of first Brillouin zone extension implies that the bandwidth achieved in metamaterials is significantly larger than the one in the structured medium chain.

From the previous measurements, we however observe that the NDBP, though slightly increasing with the surrounding medium density, is far from being on a par with the very large group index modulation. This can easily be understood considering that the NDBP = $n_g*\Delta f/f0$ and the $n_g$ modulation is more or less compensated by the bandwidth shrinking. For a tight-binding based propagation indeed, assuming that the total bandwidth $\Delta f$ is simply proportional to the linear effective bandwidth $\delta f$, the definition of the group index $n_g = c*\delta k/2\pi\delta f$ leads to NDBP ~ $c*\delta k/f_0$. In this case, the NDBP is then independent of the bandwidth of the device and is solely proportional to the wavenumber of the modes propagating in the line defect waveguide. This explains why this quantity is restricted to small and relatively constant values in more conventional composite structures, because the wave vector is constrained by the wavelength.

In our case, though the wavenumbers are the same for all waveguides, we still observe a small variation in the NDBP. This can be attributed to a slight deviation from the tight-binding model that is highlighted by the experimental determination of the dispersion relations of the wave propagating in the waveguides (see SI for protocol). Indeed, having a closer glimpse at those dispersion curves (Fig. 2e), we see that they are not purely sinusoidal, as expected in CROWs for instance, especially for the largest densities. This highlights that the propagation mechanism, while remaining principally tight binding, also takes into account a small propagative part whose contribution depends on the strength of the field confinement around the defects. It gets more important when the bandgap attenuation efficiency is decreased, leading to dispersion relations presenting some features of a polariton, which is the expected dispersion relation when resonant wires are solely coupled through a propagating wave. Beyond these propagation mechanism considerations, we observe that these curves further confirm the group index modulation through the change in the slope near the central frequency.

Despite the small increment of the NDBP while increasing the surrounding medium density, it remains relatively limited, due to the drastic bandwidth reduction. This leads to the following question: can the NDBP be further enhanced or, in other words, can the group index be largely increased while limiting the bandwidth shrinking? In the previous samples, we only tuned the group index via the coupling strength $\kappa$, i.e. via the bandwidth $\delta f$ but $a_g$, and hence $\delta k$, remained constant, and so did the NDBP, as in structured media. Contrarily to the latter however, the wavenumber $\delta k$ in metamaterials can be as well adjusted, providing a second lever that directly influences the NDBP, which is what we demonstrate in the following section.

**Enhanced delay-bandwidth products through wavenumber modulations.** To modulate the wavenumber of the modes within the waveguide, we again take advantage of metamaterials unique properties to modify the density of defects in the waveguide, namely $a_g$. Decreasing this period indeed leads to higher wavenumbers and consequently increases the group index. However, putting the defects closer also has one opposite consequence on the velocity: it facilitates the propagation through the augmentation of the coupling strength, hence decreasing the group index. As we previously demonstrated though, this coupling can as well be controlled by the surrounding medium wires' density independently of the waveguide geometrical parameters. In order to limit the impact of the waveguide density on the coupling strength, we propose to implement a scaling of the device's dimensions that amounts to simultaneously modify the medium density (as in the previous paragraph) and the waveguide lattice constant. We then manufacture two components (Fig. 4a); the reference one (Sc = 1) is a straight waveguide with the lattice constants a = 10 mm ($\lambda_0/6.5$) for the medium and $a_g$ = 6.97 mm ($\lambda_0/8.5$) for a 18 cm long waveguide. The second one (Sc = 0.5) has a and $a_g$ reduced by half and so is the waveguide length $L_g$. The spectral and temporal measurements demonstrate a striking result. While this scaling operation leads to a group delay multiplied by a factor 2.4, the transmission bandwidth reduces only by 25% (Fig. 4b,c,e,f). This value is far beyond the 50% bandwidth reduction that would be expected for an equivalent slowing strength when the coupling $\kappa$ is alone modulated, keeping the wavenumber constant (Fig. 2d). As a direct consequence of the increase of the wavenumber $\delta k$ of the slow wave modes, resulting from the geometrical scaling of the metamaterial, the NDBP is then multiplied by a factor 1.7, going from 4 to 7.5. This result completely overcomes the previous preconception of a systematic dramatic bandwidth reduction for large delay propagation in slow wave components. To complete the study, we again experimentally measure the dispersion relations and compare it with the theoretical tight-binding ones (Fig. 4d,g). As for the previous waveguides, we observe a slight deviation that can be attributed to a partly polaritonic-like behavior. This happens particularly for the Sc = 0.5 sample for which the smaller period in the waveguide may not completely prevent from direct propagative coupling. This could be avoided by decreasing the coupling between neighbors that is with higher medium density samples or larger periodicities in the waveguide.

**Delay-bandwidth products as a function of the metamaterial spatial scale**. To confirm these experimental results, we finally prove that the scaling operation can be further implemented and that the denser we scale the metamaterial samples, that is the larger the wavenumber of the modes are, the higher the delay-bandwidth products are. Because our fabrication process has a limited resolution imposed by our 3D printer, we simulate, using the software CST microwave studio, twelve of the previous devices with scaling factor ranging from Sc = 2 to Sc = 0.125. We observe the same trend as in experiment: when Sc decreases, while the normalized transmission bandwidth shrinks by a factor of 6, from 18% to 3% of the central frequency (Fig. 5a,b), the group index increases by more than two orders of magnitude (Fig. 5b). It results in NDBPs with very large values, up to 16, that seem to follow a linear dependence to a power of Sc close to -1 (Fig. 5c). It is consequently far from being limited to constant values. Again, we explain this by the fact that the NDBP value principally depends on the wavenumber of the waves propagating in the waveguide. The slight difference from the expected power -1 could be the consequence of the deviation of our dispersion relation from the pure tight-binding one that introduces a small dependence of the NDBP on the bandwidth. Note again that such a scaling operation is simply impossible to realize with other structured media since their spatial scale imposes their operating frequency, hence setting a fixed maximum value for $\delta k$.

**DISCUSSION**

Studying these line defect metamaterial waveguides have enabled evidencing three key features that can together be beneficial to both spatial and temporal control of wave propagation, especially because they exhibit certain advantages over other typical implementations of slow-wave devices. First, it allows a wave guiding with transverse dimensions that are deeply subwavelength. This stems from the fact that the hybridization bandgap in locally resonant metamaterial, that is here exploited to create defects, is rather due to the resonance of its constituents than to their spatial organization, as it is the case in structured media. The main consequence is that it leads to an extreme compactness, with dimensions actually independent of the wavelength and whose sole limitation is the resonators size, providing though that there is no direct coupling between the latter. Second, the slowing strength of these line defect waveguides is very effective, with experimental group indices as high as 227 that can be increased by further densifying the bandgap medium. In other words, the more compact it is, the slower the wave propagates, without affecting too much the operating frequency, which again is a unique prerogative of metamaterial line defect waveguides. These first two characteristics combined, deep subwavelength wave confinement and long interaction time, perfectly fulfill the two requirements for increased wave-matter interactions. Finally, using metamaterials to create very dense line defect waveguides permits to obtain unprecedented and unrestricted delay-bandwidth products. As explained in Figure 3, this comes as a less trivial consequence of the deep subwavelength scale of metamaterials allowing very large wavenumbers for the slow wave modes.

At the light of these findings, we believe that the concept we propose, line defect waveguides in metamaterials to route and slow down waves, demonstrated here in microwaves, can be of substantial interest, especially for low frequency waves. These domains indeed suffer a lack of practical solutions for both compact broadband components and wave-matter interaction enhancement platforms, despite the many applications they could benefit to. We can for instance mention microwave/atom interaction based phenomena such as NMR, widely used in medical imaging, MASERs[6,7], quantum electrodynamics implemented with microwave cavities[5], or even NV centers[8,9] and spin manipulation that are serious candidate for quantum computing. At larger scales, microwave/bulk material interaction are fundamental to observe phenomena stemming from magneto-electric or nonlinear effects for instance. Amongst potential applications are non-reciprocal devices, magnetic field sensors or data storage. From a more device-oriented point of view, in microwave to THz, compact delay lines as well as flexible wave guiding are required for all analog signals routing, radars, sensors, detection of hazardous products etc. Finally, since our concept is general, it could be easily transferred in the field of acoustic and phononics, using Helmholtz resonators[47], pillars[48] or vibrating rods[49]. Wave/matter interaction in the field of phononics are indeed equivalently important, with the development of acoustic analog signal processing using surface acoustic waves[10]. Hence, transposing to the metamaterial field concepts usually exploited with photonic crystal/CROWs, for which research has been very active in optics, we showed two things. First, the opening of new perspectives for low frequency wave manipulation (from microwave to THz, acoustics and elastic waves) and second, the evidence that the common preconception of severe bandwidth restrictions when slowing down waves is actually no fatality and is no more relevant while using metamaterials.

Naturally, the line defect waveguides suffer from some limitations. For instance, as many slow wave devices, they are dispersive (Fig. 2e, Fig. 4d,g) so that the effective operating bandwidth is in practice restrained. We can however stress that the dispersion is linear (i.e. the propagation dispersionless) over quite a wide range of frequencies near the central frequency, in agreement with the principal tight-binding contribution in the propagation mechanism. This in turn differs drastically from photonic crystal waveguides slow wave propagation occurring only on the edge of the operating bandwidth[19] or "rainbow trapping"[50,51] and slow wave approaches based on polaritonic dispersion systems as spoof plasmons, for which very large group indices only occur on the restrained flat band edge of the dispersion relation[52–55].

Moreover, if any concrete application was to be designed, attention should be taken to minimize the losses that are particularly problematic when dealing with slow waves, given the long propagation time spent in the material. In our specific microwave design, the losses, which can be attributed to the metal intrinsic properties, approximately amount to 0.1 dB/ns. The performance of the metamaterial line defect waveguides in terms of attenuation is then comparable to the best coaxial cables while largely outperforming by one order of magnitude microstrip line based delay lines (see SI). It is moreover important to stress that, on the contrary to these common microwave devices, the fair attenuation performances of our system are coupled with an extreme compactness of the waveguide. The required propagation length to reach delays of tens of nanoseconds is indeed the order of the wavelength (few centimeters) while coaxial cables would have for instance been hundreds of wavelengths long (few meters). Usually, this bulk issue is overcome by implementing microwave photonics[56] components that require modulating microwave signals forth and back to optical frequencies before delaying them in compact photonic delay lines. Though quite functional, this method undergoes several sources of substantial losses (insertion loss, conversion…) while providing typical optical delays of only few picoseconds that barely fit microwave requirements. Using metamaterial line defect waveguides then seems to be conceptually of interest to slow down microwave signals, in terms of large delays, compactness and reasonable losses.

Of course, the metal intrinsic properties prevent from transposing this specific design (metallic resonant wires) directly in optics, where material losses would severely hinder propagation. However, we can imagine that the concept itself might as well be adaptable to higher frequencies, provided the use of more suitable materials or resonators (as cold atoms for instance) taking into account the inherent specificities of optics.

**CONCLUSION**

To conclude, we demonstrated a new approach to slow down waves based on metamaterial subwavelength line defect waveguides. Owing to the subwavelength structure of metamaterials, we were able to demonstrate both very large group indices, here experimentally up to 227 and numerically up to 600, and relatively large operating bandwidths. Consequently, we measured unprecedented normalized delay-bandwidth products, more than one order a magnitude higher than prior propositions based on structured materials and that can be further increased by a simple scaling of the physical dimensions of the devices. We believe that our concept that achieves both extremely high confinement and overcomes the current limitations related to the delay-bandwidth product of slow wave devices can be of great interest in many fields of research relying on wave/matter interactions, especially for low frequency waves. It also paves the way to the design of ultra-compact components that spatially and temporally control the propagation of waves at scales independent of the wavelength. Finally, we would like to emphasize that though demonstrated here in microwave with a very specific design, our approach is very general. If it seems to suit best the requirements of low frequency regimes for now, the concept might in the future be transposed to other frequency ranges (up to THz and eventually optics) provided the use of appropriate materials. Furthermore, the idea can largely be adapted to other kind of resonators, as long as near-field interactions between unit cells can be safely neglected, or even for different type of waves such as acoustic or elastic waves.

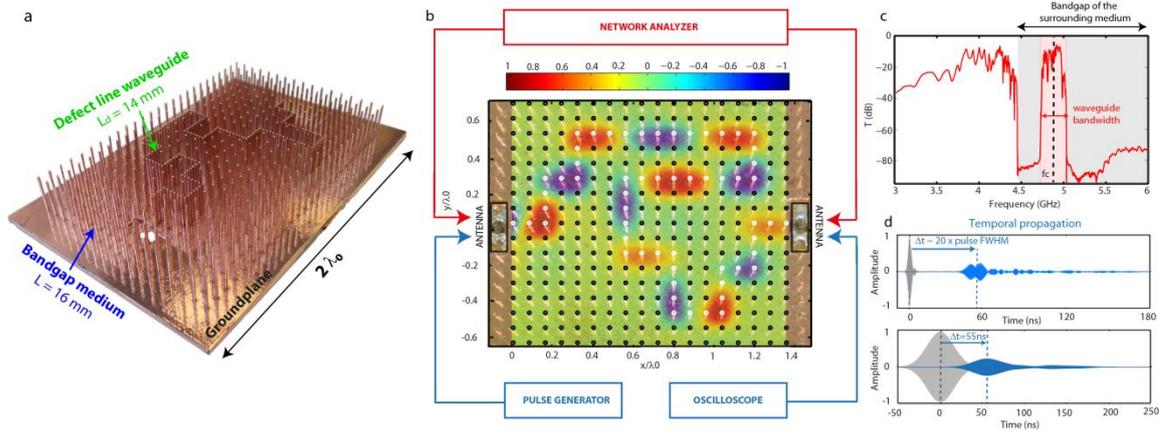

**Figure 1| Subwavelength waveguiding demonstration in a tortuous metamaterial waveguide. a**. Photograph of a 3D printed and chemically metallized wire array sitting on a ground plane comprising a tortuous defect line made out of shorter wires. To highlight the presence of the defect line, we represent in this picture a waveguide with defects separated by ag=1.5 mm, a distance three time smaller than in the sample we actually investigated the properties. **b,** Real part of the experimental E-field at f = 4.795 GHz, mapped in superimposition with a top view of the photograph of the device. Dark circles represent wires in the surrounding medium while white ones highlight the defects waveguide path. Schematic representation of the measurement set-up for the spectral (resp. temporal) properties in red (resp. blue). **c**, Spectrum of the transmission (dB). **d** (up) Temporal propagation of a short pulse centered on $f_c$ with both emitted (grey) and received pulses (blue) and (down) same for a 20 MHz wide Gaussian filtered pulse.

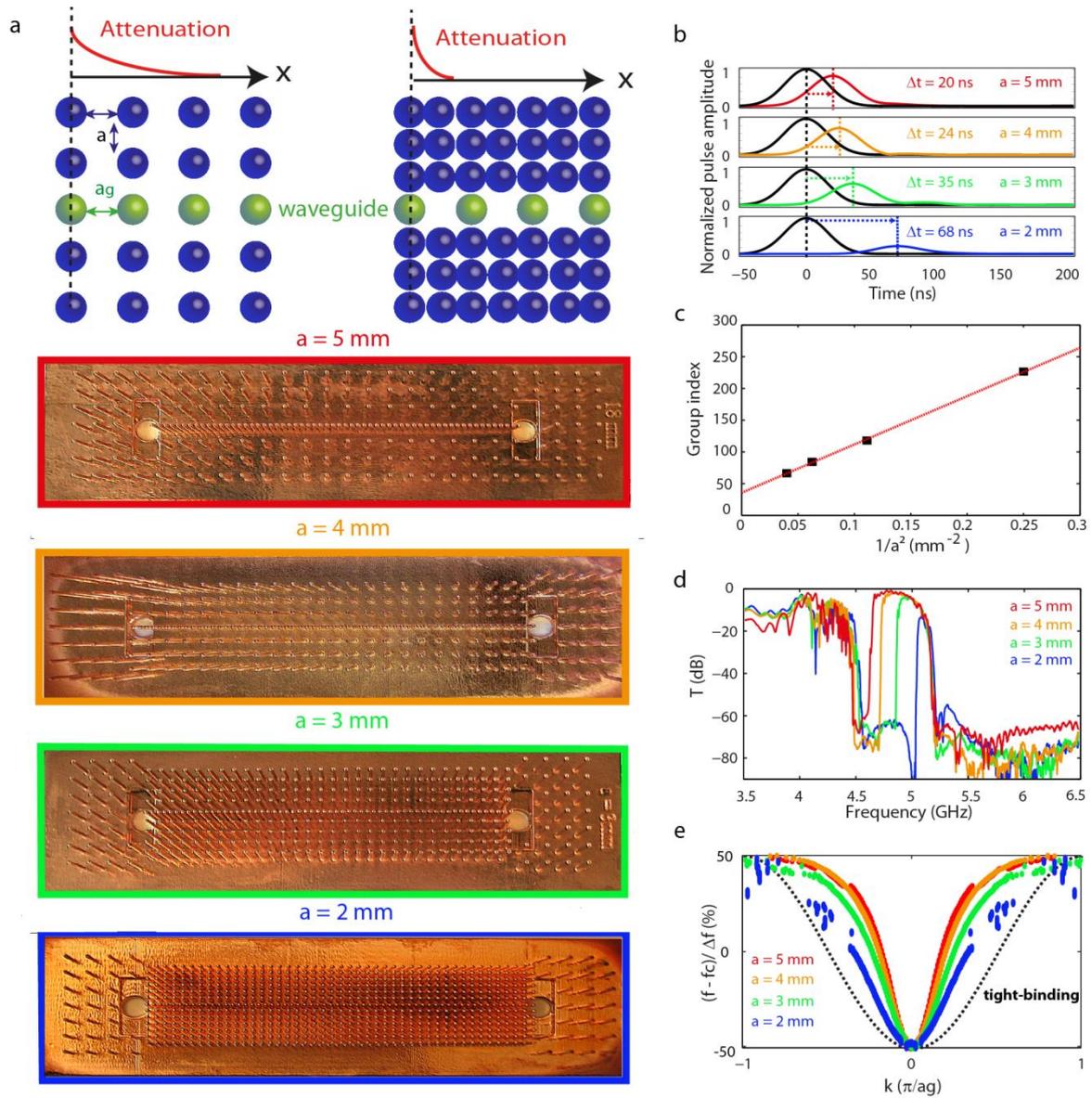

**Figure 2| Effect of the bandgap medium density on the transmission properties of the slow wave line defect waveguides. a**, Schematic representation of waveguides embedded in a bandgap medium with different densities and photographs of the four investigated devices. **b-e** Experimental results for all samples with densities ranging from a = 2 mm to a = 5 mm comprising: **b**, the envelope of the sent (black) and propagated (color) pulses at central frequency $f_c$, **c**, the group index at central frequency $f_c$ as a function of $1/a^2$, **d**, the transmission spectra (dB) and **e**, the dispersion relations displayed with normalized wavenumbers and with the frequency centered on the central frequency and normalized by the bandwidth for each waveguide.

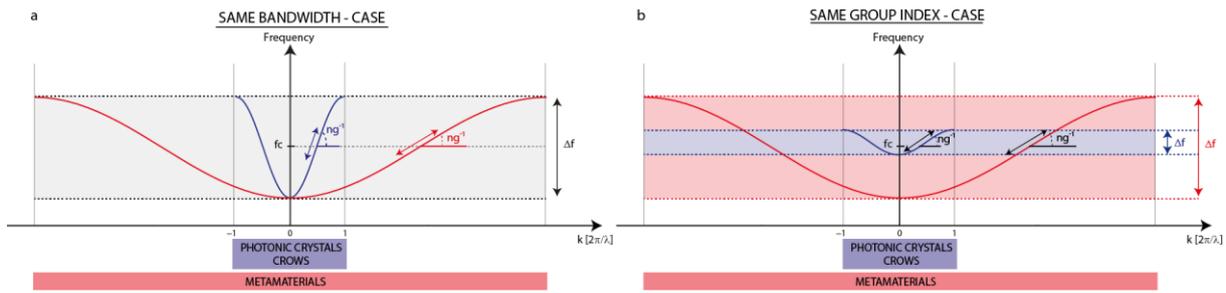

**Figure 3|Comparison of the delays and bandwidths of tight-binding dispersion relations chains in the case of coupled cavities with wavelength-scaled (photonic crystal/CROWs) and deeply subwavelength (metamaterial) periodicities.** Dispersion relation for a typical periodicity in photonic crystals coupled cavities waveguide or CROWS that is the order of the wavelength at resonance (blue) and for a subwavelength period as it is the case in metamaterials (red). (a) Comparison of the expected group index in case both materials present the same bandwidth. (b) Comparison of the expected bandwidth in case both materials present the group index around the central frequency.

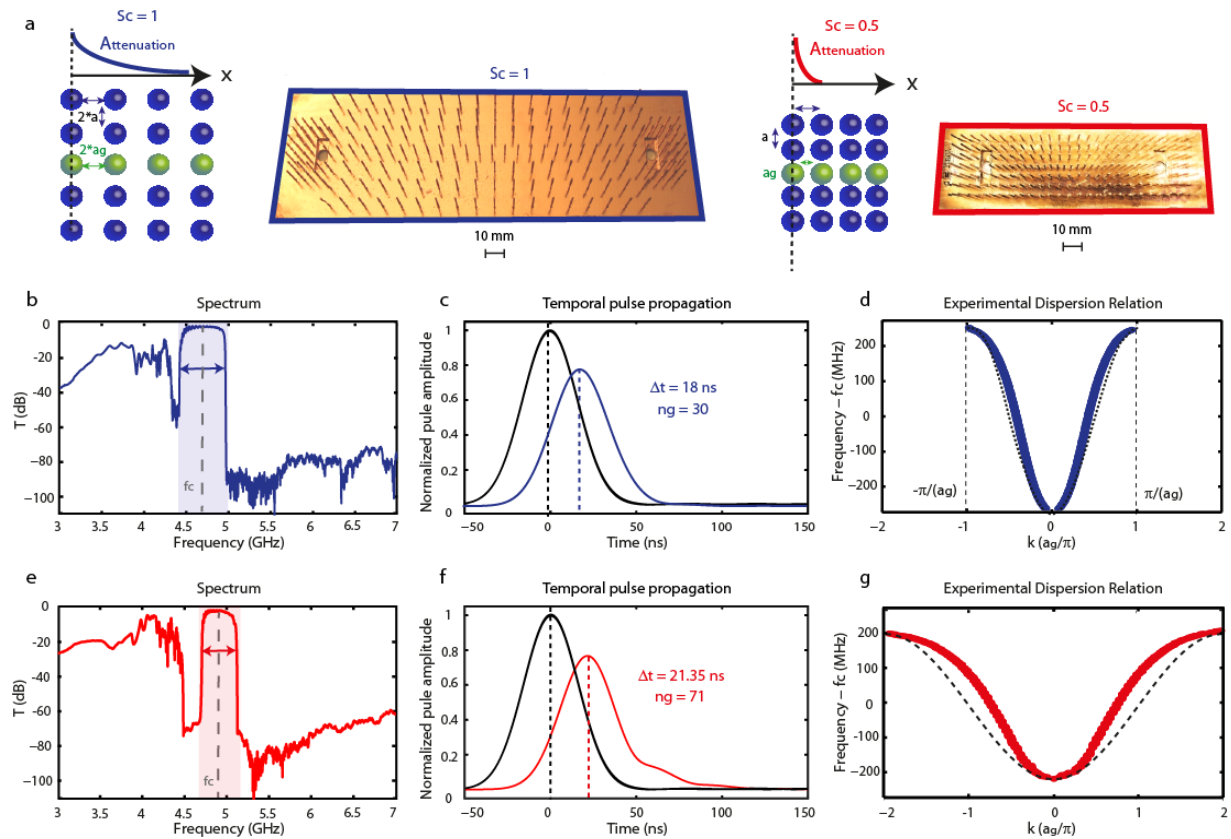

**Figure 4| Effect of the device scaling on the transmission properties of the slow wave line defect waveguides. a**, Schematic representations and pictures of the scaled samples Sc = 1 (left) and Sc = 0.5 (right). **b-d** Transmission spectrum (dB), envelope of the sent (black) and propagated (color) pulses at central frequency $f_c$ and profiles of the dispersion relations along with the theoretical tight-binding model (dotted black) for the Sc = 1 sample. **e-g** Same for the Sc = 0.5 sample. The wavenumbers in **d** and **g** are normalized by the waveguide periodicity $a_g$ in the sample Sc = 1.

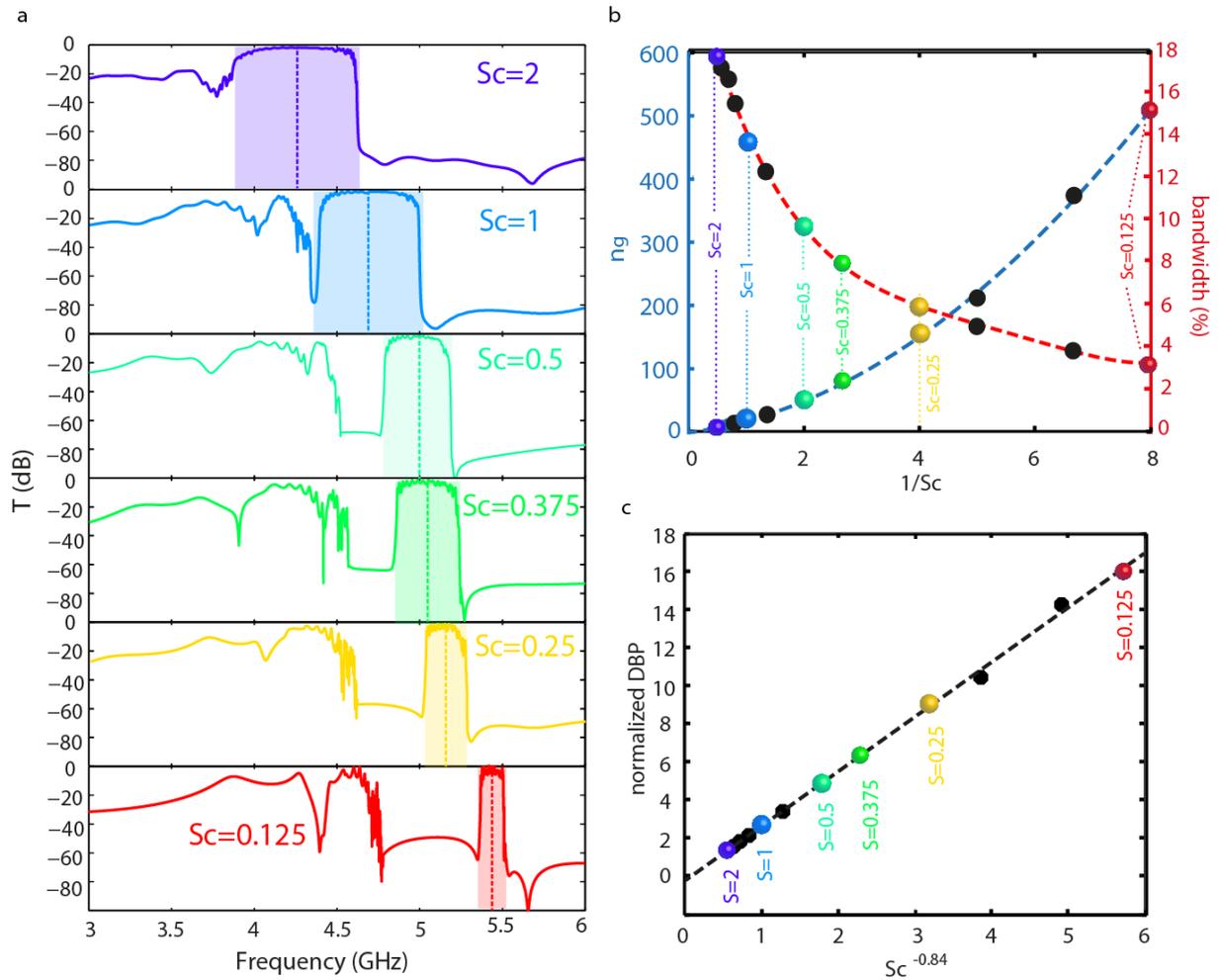

**Figure 5| Simulation of the effect of the sample scaling on the delay-bandwidth products of the slow wave line defect waveguides. a**, Simulated transmissions (dB) for six out of twelve devices with scaling factors from Sc = 2 to Sc = 0.125. **b** Evolution of the group index $n_g$ at the central frequency (y-axis left in blue, dotted blue line and black points) and the transmission bandwidth in % of the central frequency (y-axis right in red, dotted red line and black points) as a function of the inverse of the scaling factor 1/Sc. The colored points on the two curves correspond to the values of scaling factors represented in **a**. **c**. Normalized delay-bandwidth product as a function of $Sc^{-\alpha}$, where $\alpha=0.84$ gives a linear dependence. Black dotted line is the linear fit and colored points correspond to the values of scaling factors represented in **a**.